\documentclass[11pt]{article}
\usepackage{amssymb,amsmath}
\usepackage{doublespace}
\usepackage{epsf}

\begin{document}
\begin{spacing}{1.0}

\title{ON THE GENERALIZED KRAMERS PROBLEM WITH EXPONENTIAL MEMORY
FRICTION\footnote{Abbreviated Running Title: ON THE GENERALIZED KRAMERS
PROBLEM}}

\author{Katja Lindenberg\footnote{Corresponding author. Department of
Chemistry and Biochemistry 0340, UCSD, 9500 Gilman Dr., La Jolla, CA
92093-0340. Phone: (619) 534-3285. FAX: (619) 534-7244. Email:
klindenberg@ucsd.edu}\\
Department of Chemistry and Biochemistry 0340\\
and Institute for Nonlinear Science\\
University of California San Diego\\
La Jolla, California 92093-0340\\
\and
Aldo H. Romero\\
Department of Chemistry and Biochemistry 0340\\
and Department of Physics\\
University of California, San Diego\\
La Jolla, California 92093-0340\\
\and
Jos\'{e} M. Sancho\\
Institute for Nonlinear Science 0402\\
University of California San Diego\\
La Jolla, California 92093-0402\\
and\\
Departament d'Estructura i Constituents de la Mat\`{e}ria\\
Universitat de Barcelona}
 
\date{\today}
\maketitle
\end{spacing}
\begin{spacing}{1.5}

\begin{abstract}
The time-dependent transmission coefficient for the
generalized Kramers problem with exponential memory friction 
has recently been calculated by Kohen
and Tannor [D. Kohen and D. J. Tannor, J. Chem. Phys. {\bf 103}, 6013
(1995)] using a procedure based on the method of reactive flux and the
phase space distribution function. Their analysis is restricted to the
high friction regime or diffusion-limited regime.  We recently
developed a complementary theory for the low-friction
energy-diffusion-limited regime in the Markovian limit
[Sancho et al., to appear in J. Chem. Phys.; cond-mat/9806001]. Here we
generalize our method to the case of an exponential dissipative
memory kernel.  We test our results, as well as those of Kohen and
Tannor, against numerical simulations.  
\end{abstract}



\section{Introduction}
\label{intro}

The classic work of Kramers \cite{Kramers} on reaction rates
in which the effect
of the solvent was taken into account in the form of a Markovian
dissipation
and Gaussian delta-correlated fluctuations has spawned an enormous and
important literature that continues to flourish
\cite{Hanggi,Melnikov2,Pollak,Tuckerrev}).  
The enormous literature on the theoretical front since the appearance
of Kramers' seminal paper has evolved in many directions
that include more formal derivations of Kramers' own results,
extensions to larger parameter regimes and to non-Markovian
dissipation models,
generalizations to more complex potentials and to many degrees of freedom,
analysis of quantum effects, and application to specific experimental
systems. 

One recent direction, developed by Kohen and Tannor
(KT) \cite{Kohen}, deals with 
the derivation of the rate coefficient in the Kramers problem and in the
more general Grote-Hynes problem \cite{Grote}
(that is, the Kramers problem extended to a non-Markovian
dissipative memory kernel) so as to obtain not only the asymptotic rate
constant but the behavior of the rate coefficient at all times.  This
derivation is
based on the reactive flux method \cite{Hanggi,Strauba,Straubb},
thus paralleling closely and usefully
the methods used in numerical simulations of the problem.  KT
analyze in detail 
the time it takes the rate coefficient to reach its stationary
(equilibrium) value and the way in which this value is approached.
Their extensive analysis,
however, does not cover a number of parameter regimes, nor do they check
their time-dependent results against numerical simulations.  In this
paper we complement their work by extending the parameter regime of
analysis and checking their results as well as our new ones against
numerical simulations.

The non-Markovian generalization of the Kramers problem
is based on the dynamical equations \cite{Grote}
\begin{align}
\dot{q}&=\frac{p}{m} 
\notag \\ \notag\\
\dot{p}& = -\int_0^t dt'~ \Gamma(t-t') ~ p(t') - \frac{dV(q)}{dq} +f(t)
\label{generic1}
\end{align}
where $q(t)$ is the time-dependent reaction coordinate, a dot denotes a
time derivative, $\Gamma(t-t')$ is the dissipative memory kernel,
$V(q)$ is the potential energy, and $f(t)$ represents Gaussian
fluctuations that satisfy the fluctuation-dissipation
relation
\begin{equation}
\left< f(t)f(t')\right>=k_BT~\Gamma(t-t').
\end{equation}
$k_B$ is Boltzmann's constant and $T$ is the temperature.
The potential $V(q)$ is a double-well potential that is often (and here as
well) taken to be of the form
\begin{equation}
V(q)=V_0\left(\frac{q^4}{4}-\frac{q^2}{2}+\frac{1}{4}\right) =
\frac{V_0}{4}(q^2-1)^2.
\label{potential}
\end{equation}
The parameter $V_0$ can be used as the unit of energy, and
henceforth we set it equal to unity.  The barrier height is assumed
to be large compared to the temperature (i.e., $k_BT\ll 1/4$).

The problem of interest is the time-dependent rate coefficient $k(t)$
for an ensemble of particles whose positions evolve as
realizations of $q(t)$.  The coefficient $k(t)$ measures the 
mean rate of passage of the ensemble across the barrier at
$q=0$ from one well to
the other.  The asymptotic value of this crossing rate, 
$k\equiv k(\infty)$, is associated
with the rate constant of the process represented by the reaction
coordinate.  One usually focuses on the corrections to the rate obtained
from transition state theory (TST) and therefore writes
\begin{equation}
k(t)=\kappa(t) k^{TST}
\label {k}
\end{equation}
where $k^{TST}$ is the rate obtained from transition state theory for
activated crossing, which for our problem and in our units is
\cite{Hanggi,Melnikov2,we}
\begin{equation}
k^{TST}=\frac{\sqrt{2}}{\pi}e^{-1/4k_BT}.
\end{equation}
The deviations from this rate constant are
then contained in the transmission coefficient $\kappa(t)$.  
The construction of $\kappa(t)$ is discussed in subsequent sections.

\begin{figure}[htb]
\begin{center}
\leavevmode
\epsfxsize = 4.2in
\epsffile{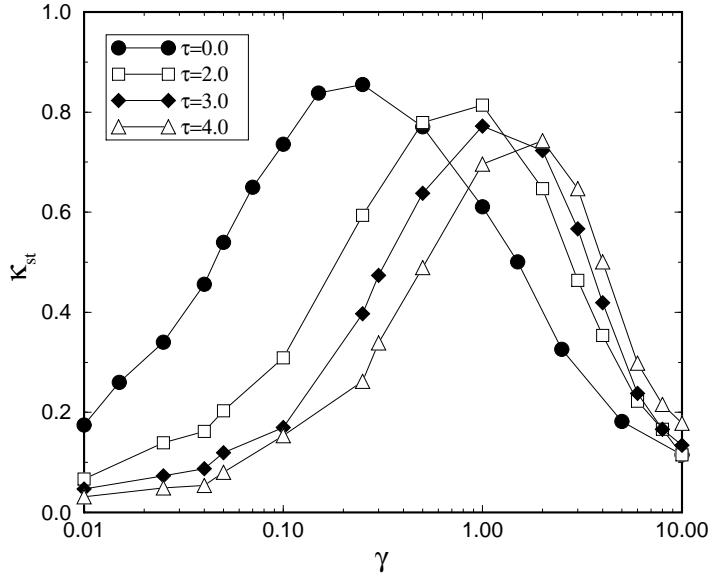}
\end{center}
\caption
{
Transmission coefficient $\kappa_{st}$ vs dissipation parameter $\gamma$
for $k_BT=0.025$ and various values of the memory correlation time $\tau$.
These results are obtained from direct simulation of Eq.~(\ref{generic1})
as described in Section~\ref{simulations}.  As is well known,
the turnover moves to higher values of the friction parameter with
increasing $\tau$ because the effective dissipation at given $\gamma$
is weaker as $\tau$ increases. Also, the maximum transmission
coefficient decreases with increasing $\tau$.
}
\label{fig2}
\end{figure}

In this paper, as do KT, we deal with the exponential memory kernel
\begin{equation}
\Gamma(t)=\frac{\gamma}{\tau}e^{-|t|/\tau}
\label{expmem}
\end{equation}
where $\gamma$ is the dissipation parameter and
$\tau$ is the correlation time.  The Kramers problem corresponds to
the limit $\tau\rightarrow 0$.  In their work, KT
derived an expression for $\kappa(t)$ in
terms of the time-dependent phase space density, assumed a particular form
for this density, and studied how $\kappa(t)$ goes to its steady state value
as the phase space density relaxes to equilibrium.  In carrying out this
program, however, they relied on approximations that are appropriate
only for the diffusion limited regime (large $\gamma$)
and therefore were able to calculate
$\kappa(t)$ only in this regime.  

In a recent paper \cite{we}
we developed a theory for the time-dependent transmission
coefficient in the energy-diffusion-limited
regime (small $\gamma$) for the original Kramers model.
Thus, to supplement the work of KT and our own work in the Markovian
regime, here we formulate the complementary theory for the time-dependent
transmission coefficient in the non-Markovian
energy-diffusion-limited regime and compare
our results with
numerical simulations in this regime.  As in the Markovian case \cite{we},
our theory captures the complex
oscillatory and stationary state behavior of $\kappa(t)$ in this
regime very well.  We also
carry out simulations of the time-dependent transmission
coefficient in the diffusion-limited regime to assess
and confirm the validity of the KT formulation.  We will show examples
of the good agreement of their results with simulations,
both in the ``non-adiabatic" regime (large $\gamma$
but $\gamma < \tau$) and in the ``caging" regime (large $\gamma$ with
$\gamma > \tau$). 

In Section~\ref{sec2} we briefly review the reactive flux formalism and
describe the numerical methods used in our simulations to implement
the formalism. Typical simulation results in various regimes are shown in
this section.  In Section~\ref{theory} we recall the theoretical
methods of analytic calculation of the transmission coefficient in the
diffusion-limited regime
(that is, KT's method) and energy-diffusion-limited regime
(our method).  The generalization of our previous Markovian results to the
non-Markovian regime is detailed in this section, as are
comparisons of these various results with numerical simulations. 
We conclude in Section~\ref{conclusion}.

\section{Reactive Flux Formalism and Simulations}
\label{sec2}

\subsection{Reactive Flux Formalism}
\label{reactflux}

Two decades ago saw the development of the reactive flux formalism for
the rate constant over a barrier \cite{Hanggi,Strauba,Straubb}. 
This formalism was important because it made possible the efficient
numerical simulation of rate constants without
having to wait the inordinately long time that it takes a particle at
the bottom of one well to climb up to the top of the barrier.  In the
reactive flux method the problem is formulated in terms of the
particles of the distribution that are sufficiently energetic
to be above the barrier at
the outset.  In this way, it is not necessary to wait for
low-energy particles (the vast majority of all the particles)
to first acquire sufficiently high energies via thermal fluctuations.  

We follow the notation of KT. The
rate constant $k(t)$ in the reactive flux formalism is
\begin{equation}
k(t)= \frac{\left< \dot{\theta}_P (q_0)\theta_P[q(q_0,
v_0,t]\right>}{\left< \theta_R(q_0)\right>}
=\frac{\left< v_0\delta(q_0)\theta_P[q(q_0,v_0,t)]\right>}{\left<
\theta_R(q_0)\right>},
\label{flux}
\end{equation}
where the top of the barrier is at position $q_0=0$, $\theta_R(x)=1$
if $x < 0$ and $0$ otherwise, and $\theta_P=1-\theta_R$.  The brackets
$\left<~\right>$ represent an average over initial equilibrium conditions.
With the particular choices made in Eq.~(\ref{flux}) one is calculating
the transition rate {\em from} the left well {\em to} the right.
KT proceed through a series of steps that finally yield 
for the transmission coefficient introduced in Eq.~(\ref{k}) the relation
\begin{equation}
\kappa(t) = \frac{m}{k_B T}\int_{-\infty}^\infty dv_0 v_0
e^{-mv_0^2/2k_BT}\chi(v_0,t)
\label{transmission}
\end{equation}
where
\begin{equation}
\chi(v_0,t) = \int_0^\infty dq \int_{-\infty}^\infty dv W(q,v,t; q_0=0,v_0)
\label{chiv}
\end{equation}
and $W(q,v,t;q_0=0,v_0)$ is the conditional phase space distribution
function that corresponds to an ensemble of particles starting at
$(q_0=0,v_0)$ at $t=0$.

The (nonequilibrium) conditional probability distribution $W$
clearly lies at the crux of the calculation: it is the distribution
associated with the generalized Langevin equation (\ref{generic1}).
In Section~\ref{theory}
we discuss the approximate solution of this problem
(which can not be solved exactly) to obtain analytic
results.  The above definitions are sufficient for the numerical
simulations. 

\subsection{Numerical Simulation Method}
\label{simulations}

The numerical solution of Eqs.~(\ref{generic1}) is performed according to
the following main steps. First, we write the problem in the entirely
equivalent form\cite{Straubc,Zwanzig}
\begin{equation}
\dot{q}=\frac{p}{m} ,\qquad
\dot{p} = - \frac{dV(q)}{dq}+z,\qquad
\dot{z} =-~\gamma~\frac{p}{\tau} -\frac{z}{\tau} +\eta(t)
\label{augmented}
\end{equation}
where
\begin{equation}
\left< \eta(t)\eta(t')\right> = \frac{2\gamma k_BT}{\tau^2}\delta(t-t').
\end{equation}
The integration of this set of equations is carried out using the
second order Heun's
algorithm \cite{Gard}, which has been tested in
different stochastic problems with very reliable results \cite{Toral94}.
A very small time step is used, ranging from
$0.001$ to $0.0001$, as in
Ref.~\cite{Strauba}.  The numerical evaluation of the
transmission coefficient $\kappa(t)$ follows the
description of Refs.~\cite{Strauba,Straubb}. We start the simulation with
$N$ particles 
(between 1000 and 4000 depending on the circumstances), all of them above
the barrier at $q=0$, half with a positive velocity
distributed according to the Boltzmann distribution in energy, which
translates to the velocity distribution $v\exp(-v^2/2 k_B T)$,
and the other half with the same distribution but with negative velocities.
The transmission coefficient is extracted from these sets of simulated
data by calculating \cite{Strauba}
\begin{equation}
\kappa(t) = \frac{N_+ (t)}{N_+(0)} - \frac{N_- (t)}{N_-(0)},
\label{extract}
\end{equation}
where $N_+(t)$ and $N_-(t)$ are the particles that started with positive
velocities and negative velocities respectively and
at time $t$ are in or over the right hand well (i.e. the particles for which
$q(t)>0$).

As discussed by Straub et al.
\cite{Straubb}, whereas the exact transmission coefficient
reaches a constant non-zero value that corresponds to the equilibrium
transition rate for the problem, the transmission coefficient calculated
using the reactive flux method flattens out but continues to decrease
with time.  If the temperature is low, this decrease is
slow and the results appear flat for rather long simulation times.
In this case the value of $\kappa(t)$ in this flat region is identified
with $\kappa_{st}$.  If the temperature is not so low, then the
decaying tail
is extrapolated back to its intersection with the vertical
axis according to the relation $ \kappa(t) \sim \kappa_{st} e^{- Kt}$
where $K$ is the decay constant of the tail.  
The value of the intersection is then identified as $\kappa_{st}$.

\subsection{Simulation Results}
\label{typical}

\begin{figure}[htb]
\begin{center}
\leavevmode
\epsfxsize = 4.2in
\epsffile{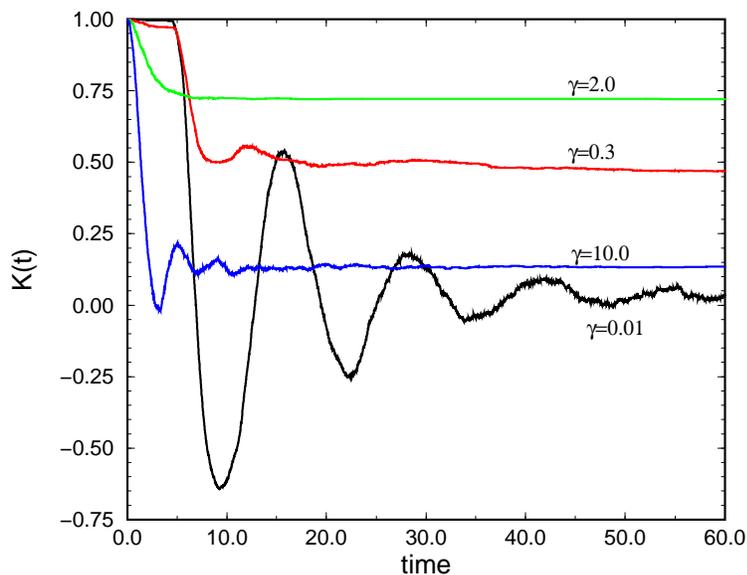}
\end{center}
\caption
{
Numerical simulation results for the transmission coefficient
$\kappa(t)$ vs time for memory correlation
time $\tau=3.0$ and temperature $k_BT=0.025$, and four values of the 
dissipation parameter $\gamma$. Diffusion-limited non-adiabatic
regime: $\gamma=2$; diffusion-limited caging regime: $\gamma=10$;
energy-diffusion-limited regime: $\gamma=0.01$; intermediate regime:
$\gamma=0.3$.
}
\label{fig3}
\end{figure}

Figure~\ref{fig3} exhibits examples of the
temporal behavior of the transmission coefficient, Eq.~(\ref{extract}),
associated with the system (\ref{generic1}),
for four typical different parameter regimes.
The input parameters of the problem are the dissipation parameter
$\gamma$, the correlation time $\tau$, and the temperature $k_BT$.
The memory correlation
time in the figure is held fixed at the value $\tau=3.0$ and the
temperature at $k_BT=0.025$. The 
$\gamma=2.0$ curve is typical of the diffusion-limited
``non-adiabatic" regime.
This is beyond the turnover value for $\tau=3.0$ in Fig.~\ref{fig2} but
still small enough to be in the non-adiabatic regime.
The transmission coefficient decays monotonically to the equilibrium
plateau. 
The curve differs from the typical behavior
in the Markovian diffusion-limited regime \cite{we,Kohen} in that
the initial decay is
Gaussian rather than exponential \cite{Kohen} (this detail is just barely
visible in the figure).  The $\gamma=10.0$ curve is typical of the regime
well beyond the turnover dissipation and well
into the ``caging regime" \cite{Kohen}.  The transmission coefficient
decays in an oscillatory fashion with an
oscillation frequency that can
be associated with an effective potential discussed later.  The
$\gamma=0.01$ curve is typical of 
very low dissipation,
well within the energy-diffusion-limited regime.  The
transmission coefficient remains at its initial value up to a time beyond
which it decays in an oscillatory manner to its equilibrium value.  As
shown later, this behavior is quantitatively captured by 
our theory. The
oscillation frequency in this case is completely different from that of the 
caging regime.  Finally, $\gamma=0.3$.
is below the turnover in Fig.~\ref{fig2} but not much below, and
the associated transmission coefficient
exhibits a mixture of features characteristic of diffusion-limited behavior
(the initial decay to an apparent plateau) and energy-diffusion-limited
behavior (the subsequent oscillatory decay to equilibrium).

\section{Theory}
\label{theory}

The conditional probability distribution 
$W(q,v,t; q_0=0,v_0)$ associated with Eq.~(\ref{generic1}) 
can not be calculated exactly for the bistable
potential (\ref{potential}).
Various approximations associated with different parameter regimes
and the resulting time-dependent transmission coefficients are
discussed in this section. 

\subsection{Diffusion-Limited Regime}
\label{difflim}
KT focus on the regime of moderate to high
dissipation, the regime first discussed in detail by Grote and Hynes
\cite{Grote}.  This is the regime to the right of the turnover in
Fig.~\ref{fig2}; the specific values of $\gamma$ included in this regime
clearly depend on the correlation time $\tau$.
To obtain an explicit expression for the 
probability distribution $W$, KT adapt the theory of Adelman
\cite{Adelman} to the assumption that the barrier is parabolic and the
wells are infinitely deep.  
With this approximation, KT
derive expressions for $\kappa(t)$ from which the Grote-Hynes
equilibrium results for the transmission coefficient are recovered
in the long-time limit.  For arbitrary times they obtain the expression
\begin{equation}
\kappa(t) = \frac{C_v(t)}{\sqrt{C_v^2(t) +\frac{mA_{11}(t)}{k_BT}}}
\label{KTresult}
\end{equation}
where 
\begin{equation}
A_{11}(t)=-~\frac{k_BT}{m}\left(C_v^2(t) - C_q^2(t)+1\right),
\end{equation}
\begin{equation}
C_q(t)=1+\int_0^t dt'~ C_v(t'),
\end{equation}
and
\begin{equation}
C_v(t) = {\cal{L}}^{-1}[s^2-1+s\hat{\Gamma}(s)]^{-1} = \sum_i c_ie^{\mu_it}.
\end{equation}
Here ${\cal{L}}^{-1}$ denotes the inverse Laplace transform,
$\hat{\Gamma}(s)$ is the Laplace transform of the memory kernel
$\Gamma(t)$, and the $\mu_i$ are the roots of
$ s^2-1+s\hat{\Gamma}(s)=0$.
For the exponential memory kernel (\ref{expmem})
$\hat{\Gamma}(s)=\gamma/(1+\tau s)$; 
the equation is then cubic so
there are three roots $\mu_i$.  The largest one of them is positive and it
determines the equilibrium transmission coefficient.  The others may be
real or complex, depending on the parameter values, and this in turn
determines whether the transmission coefficient approaches equilibrium in
an oscillatory or monotonic fashion.  KT show that in the Markovian regime
($\tau\rightarrow 0$) the decay of $\kappa(t)$ is monotonic and
exponential.  On the other hand, for $\tau$ large compared to the
time scale $\mu_1$ of the reaction (the
precise limits on $\tau$ for this to be the case are discussed in KT), 
they show that in the so-called ``non-adiabatic" regime the decay
is monotonic, whereas in the ``caging" regime
the decay is oscillatory.  This behavior is easy to understand: for a time
interval smaller than $\tau$ one can roughly approximate
$\Gamma(t-t')$ as constant.  Performing the integral in
Eq.~(\ref{generic1}) over such a time interval, one obtains an
effective potential $V_{eff}$ of the
form
\begin{equation}
V_{eff}(q)=V(q)+\frac{\gamma}{\tau}\frac{q^2}{2}.
\label{effective}
\end{equation}
If $\gamma <\tau$ (non-adiabatic regime), this is still a bistable
potential, albeit with a
smaller barrier than that of the ``bare" potential $V(q)$.  However,
if $\gamma > \tau$ (caging regime), then the effective potential
over this time interval is monostable with a frequency
$\omega_{caging}\sim (\gamma/\tau -1)^{1/2}$  about $q=0$.  

\begin{figure}[htb]
\begin{center}
\leavevmode
\epsfxsize = 4.2in
\epsffile{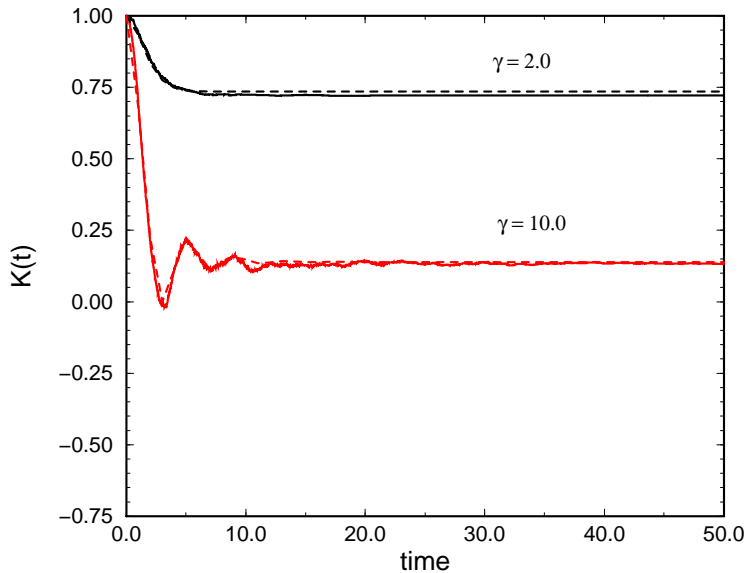}
\end{center}
\caption
{
Numerical simulation results and theoretical results for the
transmission coefficient
$\kappa(t)$ vs time for memory correlation
time $\tau=3.0$ and temperature $k_BT=0.025$, and two values of the 
dissipation parameter $\gamma$ in the diffusion-limited regime. 
Solid curves: simulation results for
$\gamma=2.0$ (diffusion-limited non-adiabatic
regime) and $\gamma=10.0$ (diffusion-limited caging regime). 
The dashed curves in each case are
calculated from the theory of Kohen and Tannor as given in
Eq.~(\ref{KTresult}).
}
\label{fig4}
\end{figure}

In Fig.~\ref{fig4} we show again two of the simulation results shown in
Fig.~\ref{fig3}, those for which the KT theory is valid, as well as the
results of Eq.~(\ref{KTresult}).  The monotonic ``non-adiabatic" decay
and the oscillatory ``caging" decay are clearly captured very well by the
theory, including the oscillation frequency in the latter (which for
these parameters is $\omega_{caging}=1.53$) and the
equilibrium values in both cases.

The KT theory is not appropriate for low dissipation parameters and indeed
it fails in that regime because it does not take into account the inertial
recrossing of the barrier that dominates this regime.  Thus, a different
approach is required in that case

\subsection{Energy-Diffusion-Limited Regime}
\label{enerlim}

When the dissipation is very slow, that is, well to the left of the
turnover in Fig.~\ref{fig2}, the dynamics of the system is dominated by the
slow variation of the energy.  Thus, a particle that starts above the
barrier may recross the barrier many times before it loses sufficient
energy to be trapped in one well or the other.  We note that
there are interesting questions concerning the role and applicability of
the reactive flux method and hence of the transmission coefficient in the
prediction of time-dependent reaction rates in the low dissipation regime
\cite{Melnikov2}.  Here we do not deal with these issues.  Rather, we take
the reactive flux method as our starting point and carry out our
analysis within that framework.  

Because of the slow energy variation it is here convenient to rewrite the
dynamical equations (\ref{generic1}) in terms of the displacement and the
{\em energy} (henceforth we set the mass $m$ equal to unity)
\begin{equation}
E=\frac{p^2}{2} +V(q)
\end{equation}
instead of the momentum \cite{Stratonovich,Ourbook}.
This simple change of variables leads to 
\begin{align}
\dot{q}&=p(t)
\label{genericq}
\\ \notag \\ 
\dot{E} &= -p(t)\int_0^t dt'~ \Gamma(t-t') ~p(t') +
f(t)p(t)
\label{genericE}
\end{align}
{\em where} $p(t)$ {\em is understood as a shorthand notation for}
$\{2[E(t)-V(q(t))]\}^{1/2}$.
This set can of course not be solved exactly either, but one can
take advantage of the fact that for slow dissipation
the temporal variation of the energy is much slower than that of
the displacement.

When dissipation is very slow, the particles that start above the barrier
lose their energy very slowly as they orbit around
at an almost constant energy during each orbit 
(recrossing the barrier many times if they start with sufficiently
high energy). One can calculate the approximate energy loss per orbit
and thus keep track of how long it takes a particle of a
given initial energy above the barrier to lose enough energy to be trapped
in one well or the other.  There is a distribution of
such particles above the barrier, and the time that it takes the ensemble
to lose enough energy to be trapped in a well is correspondingly
distributed. 

The principal ingredients in the calculation of the transmission
coefficient are: 1)
The time $t_{\varepsilon}$ that it takes an orbiting particle of energy
$\varepsilon$ above the barrier to complete a half orbit, that is, to
return to $q=0$; 2)
The energy loss in a half orbit, $\mu(\varepsilon)$, of
a particle of initial energy $\varepsilon$ above the barrier; and 3)
The distribution of times at which particles eventually become
trapped in one well or the other.

To calculate the time
that it takes a particle of energy $\varepsilon$ above the barrier
($E=1/4+\varepsilon$) to
complete one passage from $q=0$ to the edge $q_+$ of the potential
well and back to $q=0$,
we assume that the energy remains fixed during this passage.
$q_+$ is given by the solution of
$E=V(q)$, i.e., by $q_+=(1+\sqrt{4E})^{1/2}$.  Independently of the value
of $\tau$ and $\gamma$,
the time $t_\varepsilon$ for this half-orbit is then simply obtained
by integrating Eq.~(\ref{genericq}):
\begin{equation}
t_\varepsilon = 2\int_0^{q_+}
\frac{dq}{\left\{2\left[E-V(q)\right]\right\}^{1/2}}.
\end{equation}
With the potential (\ref{potential}) this can be expressed in terms of
a complete elliptic integral of the first kind.
For low temperatures the particles that determine the behavior
of the transmission coefficient are primarily those just above
the top of the barrier.  For these energies an
excellent approximation to the complete elliptic integral
yields\cite{Melnikov2,we}
\begin{equation}
t_\varepsilon \approx \ln \frac{16}{(E-\frac{1}{4})}
=\ln \frac{16}{\varepsilon}.
\label{importanttime}
\end{equation}
The corrections to Eq.~(\ref{importanttime}) are of
$O(\varepsilon\ln\varepsilon)$.

The energy of the particle of course does not in fact
remain constant as the particle orbits
around.  The calculation of $\mu(\varepsilon)$ will be postponed
for the moment -- we return to it later in this section. 
The result does
of course depend on $\gamma$, $\tau$, and temperature.  Suffice it for
now to say that to an acceptable level of approximation the energy loss
$\mu$ can be taken to be independent of $\varepsilon$.
Although this assumption is not essential, it does simplify the
calculations considerably.

The steps in the construction of the transmission coefficient from these
components have been detailed in our earlier
work\cite{we}. Here we simply display and interpret the final result:
\begin{alignat}{2}
\kappa(t) &= 1+2\sum_{n=1}^\infty (-1)^n e^{-f_n(t)/k_BT}, && \qquad 
 0<t<\ln\frac{16}{\mu},
\notag\\ \notag\\
&= 1-2e^{-\mu/k_BT} + 2\sum_{n=2}^\infty (-1)^n e^{-f_n(t)/k_BT}, && \qquad
\ln \frac{16}{\mu}<t< 2\ln \frac{16}{(2!)^{1/2}\mu},
\notag\\ \notag\\
&= 1-2e^{-\mu/k_BT} +2e^{-2\mu/k_BT}
+ 2\sum_{n=3}^\infty (-1)^n e^{-f_n(t)/k_BT}, && \qquad
2\ln \frac{16}{(2!)^{1/2}\mu}< t<
3\ln \frac{16}{(3!)^{1/3}\mu},
\label{principal}
\end{alignat}
and so on.
The function $\varepsilon=f_m(t)$ is the solution of the relation
$t=t_\varepsilon + t_{\varepsilon-\mu} +\cdots + t_{\varepsilon-(m-1)\mu}$,
which upon exponentiation turns into the $m^{th}$ order polynomial
equation
\begin{equation}
(16)^me^{-t}=\varepsilon[\varepsilon-\mu]\cdots[\varepsilon-(m-1)\mu].
\label{fm}
\end{equation}
The solution of Eq.~(\ref{fm}) can in general not be found in closed form
for $m\geq 3$.  However, an excellent approximation is
\begin{equation}
f_m(t)\approx \left[ m-(m!)^{1/m}\right]\mu +16e^{-t/m}.
\label{fma}
\end{equation}
This form is exact for $m=1$, and it is exactly correct for all $m$ at 
the upper limit of the time range that defines the trapping of the
particles whose initial energy is $\varepsilon=m\mu$.  In other words,
$f_m(t)$ as given in Eq.~(\ref{fma}) is exactly correct at the particular
time $t=t_\mu + t_{2\mu} +\cdots + t_{m\mu}$.
The asymptotic limit of Eq.~(\ref{principal}) gives for the
equilibrium transmission coefficient
\begin{equation}
\kappa_{st} \equiv \kappa(t\rightarrow\infty)=1+2\sum_{n=1}^\infty (-1)^n
e^{-n\mu/k_BT}
= \tanh \left(\frac{\mu}{2k_BT}\right).
\label{equilibrium}
\end{equation}

It is useful to examine the information contained in the various
contributions to Eq.~(\ref{principal}).
The first line contains all first
recrossings of the barrier by all particles that start with sufficient
energy above the barrier to recross is at least once
($\varepsilon > \mu$).
The function $f_1(t)$ accounts for the fact that this first recrossing
occurs at different times for particles of different initial energy,
the last ones (those closest to the barrier that do make it around)
recrossing at time $t=\ln(16/\mu)$.
The particles of higher energy recross first because they are orbiting
at higher velocity, and may be back for
their second recrossing while those of lower energy are still awaiting
their first crossing.  The time distribution of the second recrossing
is contained in $f_2(t)$.  Even while this is occurring, some particles
might already be undergoing their third recrossing, as contained in
$f_3(t)$.  There are fewer and fewer of these faster particles because of
their initial thermal distribution -- this information, too, is contained
in the exponential factors. Of course while all these recrossings are going
on, the particles are losing energy and, depending on their initial energy
and how often they have gone around, they become trapped at sequential
times as expressed in the subsequent lines of Eq.~(\ref{principal}).
In constructing these results we have averaged over a half orbit in
calculating the time for the return of a particle to the origin, and have
assumed that in the calculation of this period the energy loss of the
particle over the half orbit is negligible.  Thus, we have approximated the
evolution of the energy of a particle as determined by 
Eqs.~(\ref{genericq}) and (\ref{genericE}) with $f(t)=0$ by
simply assuming an energy loss of $\mu$ at the end of each half orbit.

The only remaining calculation is that of the energy loss $\mu$ per half
orbit. This quantity must be calculated from the generalized
Langevin equation Eq.~(\ref{genericE}).

\subsubsection{Markovian limit}

\begin{figure}[htb]
\begin{center}
\leavevmode
\epsfxsize = 3.8in
\epsffile{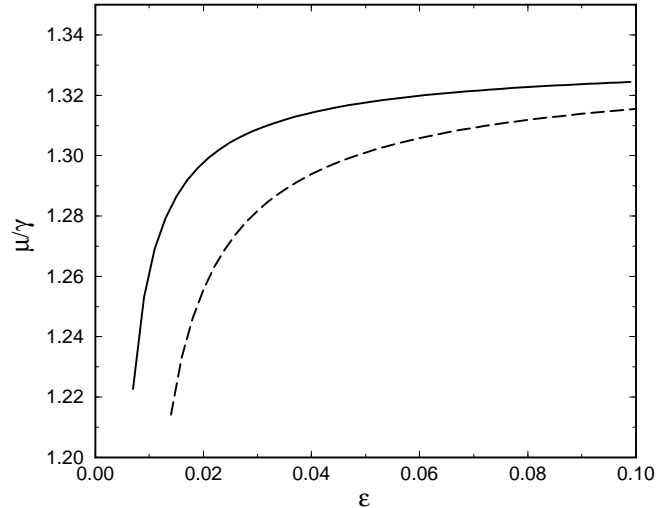}
\end{center}
\caption
{
Energy loss per half orbit as a function of energy in the Markovian
limit ($\tau\rightarrow 0$) for two values of the dissipation parameter:
$\gamma=0.005$ (dashed curve) and $\gamma=0.01$ (solid curve).
}
\label{extra}
\end{figure}

In our earlier work \cite{we} we detailed and discussed the calculation of
$\mu$ for the Markovian case.  We showed that when the fluctuations
are ignored entirely, the equation for the rate of change of the
energy, $ \dot{E}=-2\gamma[E-V(q)]$,
averaged over a half-orbit leads to the approximate equation
\begin{equation}
\dot{E}\approx -\frac{4\gamma}{3}\frac{1}{t_{E-1/4}}.
\label{2}
\end{equation}
Integration then directly leads to an implicit algebraic
equation for $\mu(\varepsilon)$ \cite{we}.
It turns out that $\mu$ does
depend on the initial energy $\varepsilon$ of the particle above the
barrier -- specifically, $\mu$ increases with increasing $\varepsilon$,
but that this dependence is mild, as seen in Fig.~\ref{extra} for two
values of the dissipation parameter.
An upper bound is the simple relation
between the energy loss per half orbit and the dissipation parameter
\begin{equation}
\mu=\frac{4}{3}\gamma.
\label{importantenergy}
\end{equation}
This result agrees with that obtained via weak-collision
arguments or small dissipation arguments by a variety of essentially
equivalent routes.  The total variation of $\mu$ with $\varepsilon$ is
of order 10\% for dissipation parameters in the energy-diffusion
limited regime, as can be seen in the figure. 

\begin{figure}[htb]
\begin{center}
\leavevmode
\epsfxsize = 4.2in
\epsffile{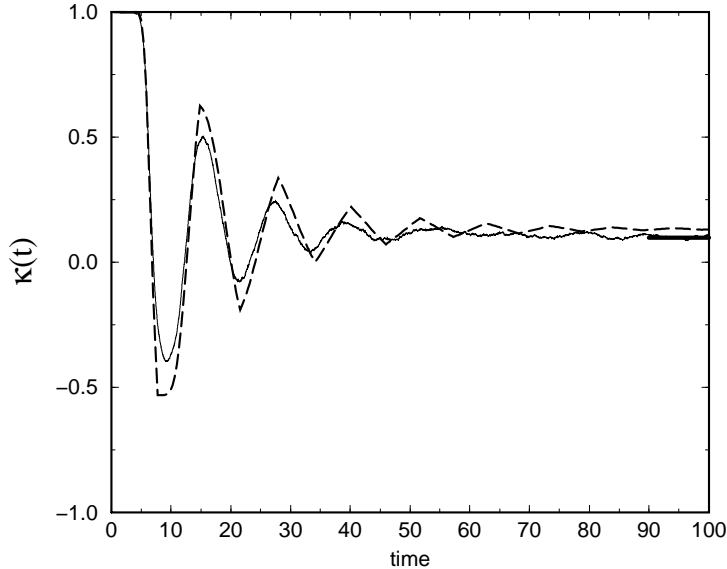}
\end{center}
\caption
{
Transmission coefficient of an ensemble of particles at temperature
$k_BT=0.025$ and dissipation parameter $\gamma=0.005$ in the Markovian
limit. 
Solid curve: simulation of $4000$ particles.  Dashed curve: our theory,
Eq.~(\ref{principal}), with the bare value $\mu=4\gamma/3=0.006667$.  
Thick line that intersects the right vertical axis: value of the
equilibrium transmission coefficient $\kappa_{st}$ obtained from 
Eq.~(\ref{equilibrium}) using the value of $\mu$ renormalized by thermal
fluctuations, which for this temperature is $\mu(T)=0.004907$. 
}
\label{fig5}
\end{figure}

Thermal effects tend to reduce $\mu$ 
since thermal fluctuations allow the particle to gain some energy as
it orbits.  We refer the reader to Ref. \cite{we} for
a discussion of thermal effects.

Figure~\ref{fig5} shows the transmission
coefficient $\kappa(t)$ as a
function of time in the Markovian limit ($\tau \rightarrow 0$) for 
a typical set of parameters in the energy-diffusion-limited regime.
The distinctive features of the time dependence are 1) the time at which
the transmission coefficient 
drops rather abruptly from its initial value of unity, and the slope
of this drop; 2) the frequencies
and amplitudes of the oscillations; and 3) the asymptotic value, identified
as the equilibrium value $\kappa_{st}$ of the transmission coefficient.
The dashed curve is the result of our theory, Eq.~(\ref{principal}), with
the bare value $\mu=4\gamma/3$ for the energy loss per half orbit. We stress
that there are no adjustable parameters in these curves.  The agreement
between the theory and simulations is clearly very good, and typical of
parameters in this regime.  Without
adjustable parameters, the theory captures each of the three 
distinctive features listed above.  Indeed, we stress that with a single
expression we are able to reproduce the temporal behavior of the system
over essentially {\em all} time scales.

The theoretical equilibrium value of the transmission coefficient
achieved by the dashed curve in Fig.~\ref{fig5} is a little
higher than the simulation result. 
We also show the equilibrium transmission coefficient obtained
when we include thermal effects \cite{we}
(thick short line that intersects the right vertical axis). 
This value of the stationary transmission coefficient agrees extremely with
the simulation result.  In \cite{we} we discuss the connection of our
equilibrium result to other predictions \cite{Melnikov3}.

\subsubsection{Non-Markovian limit}
\begin{figure}[htb]
\begin{center}
\leavevmode
\epsfxsize = 3.8in
\epsffile{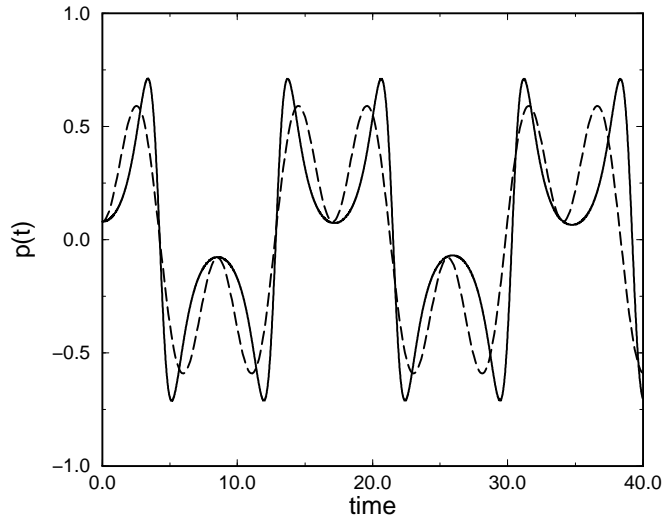}
\end{center}
\caption
{
Momentum as a function of time for a particle moving in the
potential (\ref{potential}) with energy $\varepsilon=0.003175$ above
the barrier.
Solid curve: exact result.  Dashed curve: approximation (\ref{approxp})
with (\ref{coeff}).
}
\label{fig6}
\end{figure}

To calculate $\mu$ in this regime we again begin by ignoring the
fluctuations and thus approximate the energy evolution by the first
portion of Eq.~(\ref{genericE}),
\begin{equation}
\dot{E} = -p(t)\int_0^tdt'~ \Gamma(t-t') ~ p(t'),
\label{withmemory}
\end{equation}
where $p(t)\equiv \{2[E(t)-V(q(t))]\}^{1/2}$.  To integrate this
expression over a half-orbit is complicated by the
fact that the two $p$ factors appear with different time arguments. 
One requires an analytic expression for $p(t)$ in order to carry out
this integral. It turns out that an excellent approximation to
$p(t)$ is obtained  with just two Fourier components of the form
\begin{equation}
p(\varepsilon,t)\approx
A(\varepsilon) \cos \left(\pi\frac {t}{t_\varepsilon}\right)+
B(\varepsilon) \cos \left(3\pi\frac {t}{t_\varepsilon}\right).
\label{approxp}
\end{equation}
We have explicitly indicated the energy dependence in $p$ and in the
coefficients.  The latter are fixed by imposing two conditions. One is that
the initial value of $p$ be the correct one associated with its
definition and the fact that $q(0)=0$,
\begin{equation}
p(\varepsilon,0) = A+B  =\sqrt{2\varepsilon}.
\end{equation}
The other is the equal-time average over a half orbit, already
calculated for the Markovian case [cf. Eq.~(\ref{2}]:
\begin{equation}
\frac{1}{t_\varepsilon}\int_0^{t_\varepsilon} p^2(t) dt =
\frac{A^2+B^2}{2} = \frac{4}{3t_\varepsilon}.
\label{xxx}
\end{equation}
One obtains
\begin{align}
A(\varepsilon)&=\sqrt{\frac{\varepsilon}{2}}
+\sqrt{\frac{4}{3t_\varepsilon}-\frac {\varepsilon}{2}} \notag\\
\notag\\
B(\varepsilon)&=\sqrt{\frac{\varepsilon}{2}}
-\sqrt{\frac{4}{3t_\varepsilon}-\frac {\varepsilon}{2}}.
\label{coeff}
\end{align}
Figure~\ref{fig6} is a typical example of the agreement between the exact
deterministic $p(t)$ obtained via numerical integration and the
approximation just presented.
In this figure we have chosen an energy $\varepsilon$ associated
with our subsequent discussion of the transmission coefficient.

\begin{figure}[htb]
\begin{center}
\leavevmode
\epsfxsize = 3.8in
\epsffile{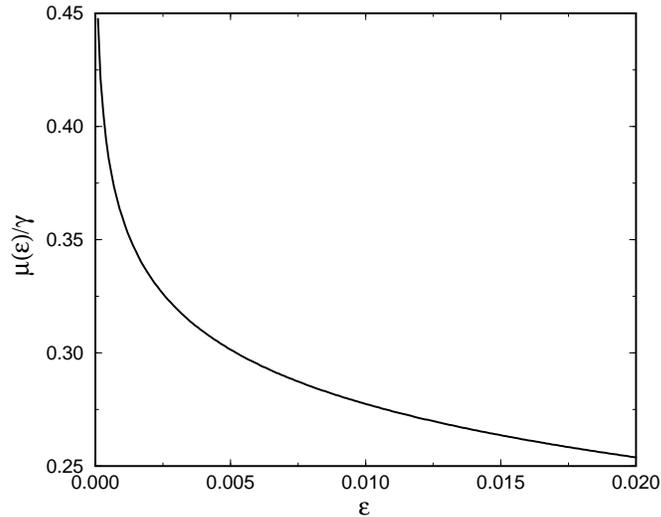}
\end{center}
\caption
{
Energy loss per half orbit as a function of energy in the non-Markovian
regime for correlation time $\tau=3.0$ and dissipation parameter
$\gamma=0.01$.
}
\label{new}
\end{figure}

With this analytic expression for $p(t)$ one can now explicitly calculate
the energy loss $\mu$ per half orbit.  Substitution of Eq.~(\ref{approxp})
into Eq.~(\ref{withmemory}), and
integration over $t$ between 0 and $t_\varepsilon$ with fixed
$\varepsilon$ immediately leads to the expression 
\begin{equation}
\frac{\mu}{\gamma}=\frac{t_\varepsilon}{2}\left(
\frac{A^2}{1+\left( \frac{\pi\tau}{t_\varepsilon}\right)^2} +
\frac{B^2}{1+\left( \frac{3\pi\tau}{t_\varepsilon}\right)^2}  \right)
-\tau\left(1+e^{-t_\varepsilon/\tau}\right) \left(
\frac{A}{1+\left( \frac{\pi\tau}{t_\varepsilon}\right)^2} +
\frac{B}{1+\left( \frac{3\pi\tau}{t_\varepsilon}\right)^2} \right)^2
\label{mu}
\end{equation}
with $A$ and $B$ given earlier.  This result is shown in Fig.~\ref{new}
for $\tau=3.0$ and $\gamma=0.01$.

The differences in Figs.~\ref{extra} and \ref{new} warrant discussion 
because of the ``opposite" behaviors of the energy loss per half orbit
with increasing energy seen in the two figures.  In the Markovian case
($\tau = 0$), the energy loss increases with increasing
$\varepsilon$. Two opposing effects contribute to this behavior. On the
one hand, the rate of energy loss is directly proportional to the
kinetic energy of the particle and, on average, the kinetic energy 
increases with increasing $\varepsilon$.  On the other hand,  the energy
loss in a half orbit increases with the time spent in this half orbit, and
this time, $t_\varepsilon$, decreases with increasing $\varepsilon$. 
These competing effects
more or less (but not entirely) compensate in the Markovian limit, and the
result is a mild increase of $\mu$ with increasing $\varepsilon$. 
The situation
becomes more complicated in the non-Markovian case, that is,
when $\tau>0$.  Now
the overwhelming factor in the $\varepsilon$ dependence of the
energy loss per half orbit is the
relation between $\tau$ and $t_\varepsilon$.  More specifically,
it is the relation between $\pi\tau$ and $t_\varepsilon$, as seen
from Eq.~(\ref{mu}).  In the regime $\pi\tau<<t_\varepsilon$ one
observes a behavior similar to the Markovian case,
that is, in this regime $\mu$ increases with $\varepsilon$
(in Eq.~(\ref{mu}) this has been
approximated by the upper bound discussed earlier, that is, by the
constant $4\gamma/3$; in any case, this
regime is not visible in Fig.~\ref{new}).
However, when $\pi\tau>t_\varepsilon$ the memory term is not fully
dissipative (cf. discussion surrounding Eq.~(\ref{effective}))
and hence leads to a smaller energy loss per
half orbit.  Since $t_\varepsilon$ decreases with increasing $\varepsilon$,
$\mu$ correspondingly decreases as well.  Thus, for finite $\tau$ the
energy loss per half orbit as a function of energy above the barrier is
expected to exhibit a maximum.  The behavior seen in Fig.~\ref{new} is
that of $\mu$ to the right of this maximum.  Note that $\tau$ would
have to be extremely small
in order to see the initial rise in $\mu$ (which in this approximation
would be a flat region).  Note also that although {\em both} Fourier
components are needed to reproduce the behavior of $p(t)$, the behavior
of $\mu(\varepsilon)$ is determined overwhelmingly by the principal
(the first) Fourier component in Eq.~(\ref{approxp}).

From the result Eq.~(\ref{mu}) we now wish to extract a specific (constant)
value of $\mu$ to use in the calculation of the transmission coefficient.
This is more difficult in the non-Markovian case than in the Markovian
case because the variation
of $\mu$ with $\varepsilon$ is much more pronounced.  However, one should
remember that at low temperatures the population above the barrier is small
and the principal contributors to the transmission coefficient are the
particles near the barrier.  Indeed, in the Markovian case we see from
Fig.~\ref{fig5} that there are about five or six oscillations before
the transmission coefficient settles down, indicating the participation
of particles in a layer of width $5\mu$ or so above the barrier.  
The variation of $\mu$ over a comparable range in Fig.~\ref{new} is
actually rather small, that is, only a small portion of the range shown
in the figure is actually important for the transmission coefficient. 
Thus, for instance, for $\tau=3$ we find that the value of
$\mu$ for those particles that half-orbit once before becoming
trapped (found from Eq.~(\ref{mu}) by setting $\varepsilon=\mu$ and solving
for $\mu$) is $\mu/\gamma=0.3175$.  For $\gamma=0.01$
this corresponds to $\mu=0.003175$.  The
variation of $\mu$ between $\varepsilon=0.003175$ and, say, five times
this value, is about 15\%.  Thus, if we take $\mu=0.003175$ we might
overestimate $\mu$ by an amount only somewhat greater than that
of the Markovian
calculation, and expect the result to be comparably satisfactory.

\begin{figure}[htb]
\begin{center}
\leavevmode
\epsfxsize = 4.2in
\epsffile{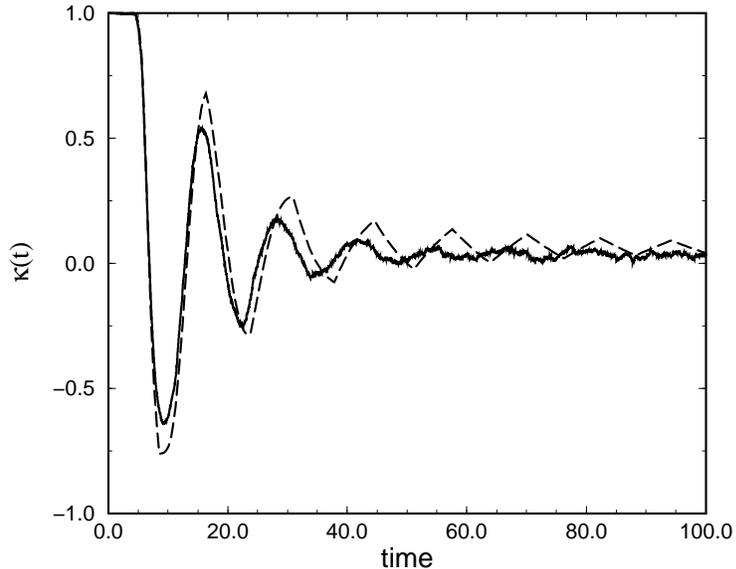}
\end{center}
\caption
{
Transmission coefficient of an ensemble of particles at temperature
$k_BT=0.025$, dissipation parameter $\gamma=0.01$, and dissipative
memory time $\tau=3.0$. 
Solid curve: simulation of $4000$ particles.  Dashed curve: our theory,
Eq.~(\ref{principal}) with $\mu=0.003175$.
}
\label{finale}
\end{figure}

In Fig.~\ref{finale} we show the simulation results as well as our
theoretical result for a typical set of parameters.  The simulation 
curve is one of those shown in Fig.~\ref{fig3}.  As in the Markovian
case, the theory captures the distinctive features mentioned
in the Markovian case: the time of the first abrupt drop of the
transmission
coefficient, the frequencies and amplitudes of the oscillations, and the
asymptotic value.  The theoretical overestimation of $\mu$ in the
later oscillations is discernible (as it is in Fig.~\ref{fig5}), and
the calculated equilibrium transmission
coefficient is also perhaps a bit high (as it is in Fig.~\ref{fig5}). 
However, we understand the causes of these small effects and know how to 
correct the theory for them if needed.   Also, the thermal corrections
would bring the equilibrium transmission coefficient into even closer
agreement with the simulation results, as in the Markovian case.
Altogether, the theory is clearly very good in its prediction of
the complex time-dependent and equilibrium behavior of
the transmission coefficient in the non-Markovian regime for the case
of an exponential memory kernel. 

\section{Conclusions}
\label{conclusion}

In this paper we have analyzed the time-dependent transmission coefficient
for the generalized Kramers problem, that is, for a particle in a bistable
potential described by a generalized Langevin equation. 
The transmission coefficient is associated with the rate of transitions
of the particle from one well to the other.  In this presentation the
dissipative memory kernel has been taken to be exponential in time.
In the diffusion-limited
regime we have confirmed the predictions of Kohen and Tannor \cite{Kohen}
via numerical simulations, and have shown their predictions to be accurate
both in the non-adiabatic regime and in the caging regime. In the
energy-diffusion-limited regime we have generalized our previous theory for
the Markovian limit \cite{we} to the non-Markovian case and have shown
good agreement with numerical simulations in this regime as well.  One can
therefore say that there is now a fairly complete understanding of the
time-dependent transmission coefficient for the generalized Kramers problem
with an exponential memory kernel in the limiting regimes of strong
dissipation and weak dissipation.

These remarks cover three of the four regimes represented by the
simulations in Fig.~\ref{fig3}, that is, the $\gamma=2.0$, $\gamma=10$, and
$\gamma=0.01$ curves.  The fourth curve, the one for $\gamma=0.3$,
displays a distinct behavior that has not been discussed in this paper
but that will be discussed elsewhere \cite{we2}.
This represents an ``intermediate" dissipation and is thus associated
with a regime in which elements appropriate to the diffusion-limited
regime and also to the energy-diffusion-limited regime in
combination play an important role.  

If one were to apply (misadvisedly) the KT theory to the left of the
turnover in Fig.~\ref{fig2}, one would find an apparent settling of the
transmission coefficient at the values that in reality represent only the
short-time plateau of $\kappa(t)$ seen in Fig.~\ref{fig3} before
the first abrupt drop.  Thus,
for $\gamma=0.01$ the transmission coefficient would settle at nearly unity
and for $\gamma=0.3$ it would settle at around 0.96.   
The actual abrupt drops seen in Fig.~\ref{fig3} following these initial
plateaus come about because of the particles that have not been trapped
after one half orbit and that at times $t_\varepsilon$ cross the barrier to
move over the left well.  The diffusion-limited theory of course does not
capture this recrossing and hence levels off.  This leveling off has been
referred to as an example of a ``false plateau" that one must be careful of
in interpreting	simulation data.  Perhaps more interesting ``false plateau"
manifestations occur as a result of memory kernels that are not exponential
and that can lead to bottlenecks of various kinds in the relaxation process
\cite{Tucker}.  In general, one expects that different memory kernels
might lead to behaviors not seen in the particular exponential
model considered in this paper.  Our analysis of other memory kernels
and the resultant exploration of interesting new dynamical manifestations
of these differences will also be presented elsewhere \cite{we2}.

\section*{Acknowledgment}
This work was done during a sabbatical leave of J.M.S. at the University
of California, San Diego granted by the Direcci\'o General de Recerca de
la Generalitat de Catalunya (Gaspar de Portol\'a program).  
The work was supported in part by the U.S. Department of Energy under
Grant No. DE-FG03-86ER13606, and in part by the Comisi\'on
Interministerial de Ciencia y Tecnolog\'ia (Spain)
Project No. DGICYT PB96-0241.

\newpage
\setcounter{figure}{0}
\begin{figure}
\caption
{
Transmission coefficient $\kappa_{st}$ vs dissipation parameter $\gamma$
for $k_BT=0.025$ and various values of the memory correlation time $\tau$.
These results are obtained from direct simulation of Eq.~(\ref{generic1})
as described in Section~\ref{simulations}.  As is well known,
the turnover moves to higher values of the friction parameter with
increasing $\tau$ because the effective dissipation at given $\gamma$
is weaker as $\tau$ increases. Also, the maximum transmission
coefficient decreases with increasing $\tau$.
}
\end{figure}

\begin{figure}
\caption
{
Numerical simulation results for the transmission coefficient
$\kappa(t)$ vs time for memory correlation
time $\tau=3.0$ and temperature $k_BT=0.025$, and four values of the 
dissipation parameter $\gamma$. Diffusion-limited non-adiabatic
regime: $\gamma=2$; diffusion-limited caging regime: $\gamma=10$;
energy-diffusion-limited regime: $\gamma=0.01$; intermediate regime:
$\gamma=0.3$.
}
\end{figure}

\begin{figure}
\caption
{
Numerical simulation results and theoretical results for the
transmission coefficient
$\kappa(t)$ vs time for memory correlation
time $\tau=3.0$ and temperature $k_BT=0.025$, and two values of the 
dissipation parameter $\gamma$ in the diffusion-limited regime. 
Solid curves: simulation results for
$\gamma=2.0$ (diffusion-limited non-adiabatic
regime) and $\gamma=10.0$ (diffusion-limited caging regime). 
The dashed curves in each case are
calculated from the theory of Kohen and Tannor as given in
Eq.~(\ref{KTresult}).
}
\end{figure}

\begin{figure}
\caption
{
Energy loss per half orbit as a function of energy in the Markovian
limit ($\tau\rightarrow 0$) for two values of the dissipation parameter:
$\gamma=0.005$ (dashed curve) and $\gamma=0.01$ (solid curve).
}
\end{figure}

\begin{figure}
\caption
{
Transmission coefficient of an ensemble of particles at temperature
$k_BT=0.025$ and dissipation parameter $\gamma=0.005$ in the Markovian
limit. 
Solid curve: simulation of $4000$ particles.  Dashed curve: our theory,
Eq.~(\ref{principal}), with the bare value $\mu=4\gamma/3=0.006667$.  
Thick line that intersects the right vertical axis: value of the
equilibrium transmission coefficient $\kappa_{st}$ obtained from 
Eq.~(\ref{equilibrium}) using the value of $\mu$ renormalized by thermal
fluctuations, which for this temperature is $\mu(T)=0.004907$. 
}
\end{figure}

\begin{figure}
\caption
{
Momentum as a function of time for a particle moving in the
potential (\ref{potential}) with energy $\varepsilon=0.003175$ above
the barrier.
Solid curve: exact result.  Dashed curve: approximation (\ref{approxp})
with (\ref{coeff}).
}
\end{figure}

\begin{figure}
\caption
{
Energy loss per half orbit as a function of energy in the Markovian
limit ($\tau\rightarrow 0$) for two values of the dissipation parameter:
$\gamma=0.005$ (dashed curve) and $\gamma=0.01$ (solid curve).
}
\end{figure}

\begin{figure}
\caption
{
Transmission coefficient of an ensemble of particles at temperature
$k_BT=0.025$, dissipation parameter $\gamma=0.01$, and dissipative
memory time $\tau=3.0$. 
Solid curve: simulation of $4000$ particles.  Dashed curve: our theory,
Eq.~(\ref{principal}) with $\mu=0.003175$.
}
\end{figure}

\end{spacing}

\end{document}